\documentstyle[12pt,a4]{article}

\begin{document}
\baselineskip 4ex
\vspace*{-2.5cm}
\hfill{\bf TSL/ISV-97-0188}
\vspace{1.0cm}
\begin{center}
{\Large{\bf Bound and Unbound Wave Functions at Short Distances}}\\[7ex]
{\large G{\"{o}}ran F\"{a}ldt}$^{a}$\\[1ex]
{\normalsize Division of Nuclear Physics, Box 535, 751 21 Uppsala, Sweden}
\\[3ex]
{\large Colin Wilkin}$^{b}$\\[1ex]
{\normalsize University College London, London, WC1E 6BT, UK}\\[3ex]
\today\\[4ex]
\end{center}

\baselineskip 4ex
\begin{abstract}
There exists a simple relationship between a quantum-mechanical bound-\-state 
wave function and that of nearby scattering states, when the scattering energy
is extrapolated to that of the bound state. This relationship is demonstrated 
numerically for the case of a spherical well potential and analytically for 
this and other soluble potentials. Provided that the potential is of finite 
range and that the binding is weak, the theorem gives a useful approximation 
for the short-distance behaviour of the scattering wave functions. The
connection between bound and scattering-state perturbation theory is 
established in this limit.
\end{abstract}
\vspace{1cm}

\noindent
{\large\bf I.\ \ Introduction}\\

It has recently been shown there is a simple relationship between the
normalisations of the scattering and bound-state wave functions when the
scattering energy is continued to that of the bound state$^{1}$. That the
two wave functions become mutually proportional in this limit is known from the
standard text books$^{2,3}$, but it was unexpected to find
that the relative normalisation depended purely upon the binding energy. Though
the theorem is valid for arbitrary angular momentum, we wish here to illustrate
some of the results for $S$-wave scattering. Provided the range of the
potential is short and the binding is light, it will be seen that the 
extrapolation theorem actually gives useful approximations for scattering 
wave functions at low energies and short distances.

The $S$-wave Schr\"{o}dinger equation for the interaction of a particle of 
mass $m$ in a potential $V(r)$ may be written as
\begin{eqnarray}
u_{\alpha}''(r) -\alpha^2u_{\alpha}(r)&=&U(r)\,u_{\alpha}(r) \nonumber \\
v''(k,r)+k^2v(k,r)&=&U(r)\,v(k,r) \, ,\label{1_1}
\end{eqnarray}
where $U(r)=2mV(r)/\hbar^2$ and the binding and scattering energies are
$-\hbar^2\alpha^2/2m$ and $\hbar^2k^2/2m$ respectively. The corresponding
radial wave functions, $u_{\alpha}(r)$ and $v(k,r)$, vanish at $r=0$ but 
satisfy contrasting boundary conditions at large distances. For a potential 
of finite range the bound-state wave function behaves asymptotically like 
\begin{equation}
\label{1_1a}
u_{\alpha}(r)\approx N_{as}\,\mbox{\rm e}^{-\alpha r}
\end{equation}
and is normalised by the integral condition
\begin{equation}
\label{1_2}
\int_0^{\infty} [u_{\alpha}(r)]^2\,dr =1\:.
\end{equation}
On the other hand the scattering solution normalisation is determined 
by its asymptotic behaviour and, if we take {\it real} boundary conditions 
(standing waves), then
\begin{equation}
\label{1_3}
v(k,r)\longrightarrow\frac{1}{k}\sin (kr+\delta(k))
= \frac{1}{2ik}\left(\mbox{\rm e}^{i\delta(k)}\mbox{\rm e}^{ikr}-
\mbox{\rm e}^{-i\delta(k)}\mbox{\rm e}^{-ikr}\right)
\end{equation}
as $r\to\infty$, where $\delta(k)$ is the $S$-wave phase shift at wave number 
$k$.

In order that $v(k,r)$ might become an acceptable localisable wave function as 
$k\to i\alpha$, for which the normalisation condition of
Eq.(\ref{1_2}) would hold, the S-matrix must have a pole at the
position of the bound
state and can therefore be written in this region as 
\begin{equation}
\label{1_4}
S(k) = \mbox{\rm e}^{2i\delta(k)} 
=N^2(k)\,\left(\frac{\alpha-ik}{\alpha+ik}\right)
\ \ \Rightarrow\ \ \mbox{\rm e}^{i\delta(k)} = 
N(k)\,\frac{(\alpha-ik)\ \ \ }{(\alpha^2+k^2)^{1/2}} 
\:,
\end{equation}
where $N(k)$ is well behaved in the vicinity of $k=i\alpha$.

At the bound-state pole the second term in Eq.(\ref{1_3}) is eliminated 
to leave the desired asymptotic behaviour 
\begin{equation}
\label{1_5}
v(k,r) \approx -\frac{N(i\alpha)\ \ }{(\alpha^2+k^2)^{1/2}}\:
\mbox{\rm e}^{-\alpha r}\:.
\end{equation}

Though this does not establish the relative normalisation of this and
the bound-state wave function $u_{\alpha}(r)$, 
the presence of the square root in the
denominator of Eq.(\ref{1_5}) is indicative and this factor is
contained within the extrapolation theorem$^{1}$, which for potentials
of finite range yields
\begin{equation}
\label{1_6}
\lim_{k\to i\alpha}
\left\{\left[2\alpha(\alpha^2+k^2)\right]^{1/2}\, v(k,r) \right\}= -
 u_{\alpha}(r) \:. 
\end{equation}
It should be noted that this relationship is {\it independent} of the
shape of the potential and is determined purely by the binding energy.
It is now easy to see from Eqs.(\ref{1_5}) and (\ref{1_6}) that, in
terms of the asymptotic normalisation constant $N_{as}$ of
Eq.(\ref{1_1a}), the residue at the pole of the $S$-matrix in 
Eq.(\ref{1_4}) equals $-iN_{as}^2$.

As will be seen from the numerical examples given for the spherical well
potential discussed in \S2, the above extrapolation theorem actually provides
useful approximations for scattering wave functions in terms of that of a 
nearby bound state. Provided that the binding is weak, and that neither 
$k$ nor $r$ are too large, 
\begin{equation}
\label{1_7}
v(k,r) \approx - \left[2\alpha(\alpha^2+k^2)\right]^{-1/2}\, u_{\alpha}(r)\:.
\end{equation}
The ratio of the scattering to the bound-state wave functions is studied
analytically for several exactly soluble potentials in \S3, where more precise
statements are made on the range of validity of the approximations in 
Eq.(\ref{1_7}) and general trends in the deviations noted. One feature seen for
the spherical well potential is that there seems to be a region of $r$ where 
all the wave functions, scaled by the square root factor as in Eq.(\ref{1_7}), 
come together. A simple understanding of this cross-over effect is also
attempted in \S3. 

The low energy proton-proton system is not quite bound but there is 
a virtual state with small but negative $\alpha$ situated close to physical
energies. Though in such cases there is no bound-state wave function to set
the scale, the energy dependence of the scattering wave functions is well 
approximated by the square-root factor as in Eq.(\ref{1_7}). This behaviour is
clearly demonstrated by the spherical well and it is seen from these examples
that it is the nearest singularity which is dominant.

Though the long range of the Coulomb potential violates the assumptions made
when proving the extrapolation theorem of Eq.(\ref{1_6}), it nevertheless 
remains valid for {\it all} the Coulomb bound states, as shown in \S4. 
However the accumulation of bound states in the vicinity of zero energy means
that Eq.(\ref{1_7}) is of no use in representing the scattering functions even
at low energies. There is no nearby {\it dominant} pole.

Eq.(\ref{1_7}) is of great help in estimating the effects of final state
interactions in nuclear processes at large momentum transfers, where the
behaviour of the wave function at small distances is important$^{4,5,6,7}$.
For example it allows one to estimate the cross section for $pp\to pn\pi^+$ in
terms of that for $pp\to d\pi^+$, where the final neutron-proton triplet 
pair are fused to become a deuteron. Another simple example discussed 
in \S5 involves the relationship between perturbation theory applied to 
bound and scattering state problems. A consistent picture is found when 
using Eq.(\ref{1_7}) in the presence of a lightly bound state. A summary 
of our main conclusions is to be found in \S6.\\[3ex]

\noindent
{\large\bf II.\ \ Numerical investigation for the spherical well potential}\\

As a first illustration of our ideas, consider the spherical well potential
\begin{equation}
\label{2_1}
U(r)= \frac{2m}{\hbar^2}\,V(r)=\left\{
\begin{array}{rcl}
-U_0 &,&r<a\,\,,\\
0&,&r\ge a\,\,. \end{array}\right.
\end{equation}

For $r\ge a$ the scattering and bound-state wave functions are proportional to
$\sin(kr+\delta)$ and e$^{-\alpha r}$ respectively, whereas inside the well 
they are rather $\sin(\kappa r)$, where $\kappa^2=U_0-\alpha^2$ or
$\kappa^2=U_0+k^2$. It is trivial to match boundary conditions at $r=a$ to
obtain an implicit equation for the energy eigenvalues and hence the wave 
functions$^{2}$.

The approximation of Eq.(\ref{1_7}) is expected to work best for a loosely 
bound state, and we start by treating the case where there is just a 
single $1s$
state. Taking units where the potential radius $a=1$, a potential strength of
$U_0=2.8$ yields $\alpha = 0.159$ and the normalised full bound-state wave 
function $\psi_{\alpha}(r)=u_{\alpha}(r)/r$ is plotted in fig.~1.1. Also shown
there are the corresponding scattering wave functions, modified by the 
square-root factor of Eq.(\ref{1_7}),
\begin{equation}
\label{2_2}
\tilde{\psi}(k,r) = -\left[2|\alpha|(\alpha^2+k^2)\right]^{1/2}\:v(k,r)/r\:,
\end{equation}
at wave numbers of $k=0.1$, $0.2$, $0.5$, and $1.0$.

At $r=0$ the agreement between the various functions is very good for low 
values of $k$, though the scattering ones lie above that of the bound state 
and steadily increase with $k$. Though the energy dependence of the functions 
in this region seems to be roughly $\kappa^{1/2}$, too much should not 
be read into this. The scattering functions must of course be
orthogonal to the bound state and this implies that they start to oscillate
before the bound state has died out. This is seen clearly in the figure, as is
the fact that the oscillations set in earlier for higher $k$-values. What is
less evident is why all the curves seem to come close together for $r\approx
0.66$. This has the practical effect however of limiting the 
deviations between the scattering and bound-state functions over the 
range of the potential.

If the potential strength is increased in order to generate a deeply bound $1s$
state in addition to a $2s$ state with the same binding as in fig.~1.1, 
then the latter has the node demonstrated in fig.~1.2. The scattering 
functions defined by
Eq.(\ref{2_1}) still fall very close to that of the bound state with the 
crucial difference that they now lie below for small $r$. A cross-over region 
still exists but at higher values of $r$.

Fig.~1.3 shows that if the attraction is reduced a little such that the 
$2s$ state becomes a virtual one with negative $\alpha$ ($\alpha = -0.159$) 
then, as expected, the scattering wave functions still retain their 
dominantly $2s$ character, though there is then no bound state to set the 
scale and $r$-dependence. The wave function of such a virtual state 
increases like
e$^{+|\alpha|r}$ at large distances and hence is non-normalisable. It is
therefore clear that it is the nearest singularity of the scattering amplitude
which governs the behaviour of the wave function at low $k$. 
Though Eq.(\ref{1_6}) is valid for
{\it all} the bound states of a problem when extrapolating to the pole,
Eq.(\ref{1_7}) is only a useful representation for a nearby singularity. An
important practical case of this kind is the proton-proton system where the
(just) unbound virtual state dominates the low energy wave function and allows
us to derive the energy dependence of the $pp\to pp\pi^0$ 
cross section near threshold$^{5}$.\newpage

\noindent
{\large\bf III.\ \ Analytic expressions for soluble potentials}\\

In addition to the spherical well potential, 
there are several other potentials for which exact solutions 
to the Schr{\"{o}}dinger equation can be obtained. In such
cases it is interesting to study the deviations from the 
extrapolation theorem analytically in going away from the bound-state 
pole at $k=i\alpha$ to see the dependence upon the parameters. 
Define therefore the ratio
\begin{equation}
\label{3_1}
R(k,r) = -\left[2\alpha (\alpha^2+k^2)\right]^{1/2}\: v(k,r)/u_{\alpha}(r)
\end{equation}

In order to satisfy the extrapolation theorem $R(i\alpha,r)=1$ and,
provided that the potential is of finite range, the ratio function can be
expanded as a power series in $\alpha^2+k^2$.
\begin{equation}
\label{3_2}
R(k,r)=1 + \sum_{n=1}^{\infty} R_{n}(r)\,(\alpha^2+k^2)^n\:.
\end{equation}
In general this will have a finite radius of convergence since
$R(k,r)$ diverges at the position of any other bound state.\\

\noindent
{\large\bf A.\ \ The Yamaguchi potential}

The strength $\lambda$ of the non-local but separable Yamaguchi potential,
\begin{equation}
\label{3_3}
V(r,r') = -\lambda\,\left(\frac{\mbox{\rm e}^{-\beta r}}{r}\right)\,
\left(\frac{\mbox{\rm e}^{-\beta r'}}{r'}\right)\:,
\end{equation}
can be adjusted to give a solitary bound state at $E=-\hbar^2\alpha^2/2m$.

A straightforward calculation then shows that to second order in $r$
\begin{equation}
\label{3_4}
R_1(r) = \frac{3}{8\beta(\beta+\alpha)}+
\frac{1}{4\beta}\,r-\frac{(\beta-\alpha)}{8\beta}\,r^2+ 0\!\left(r^3\right)\:.
\end{equation}
\vspace{2mm}

\noindent
{\large\bf B.\ \ The Bargmann potential}

The Bargmann potential defined by
\begin{equation}
\label{3_5}
U(r)= \frac{2m}{\hbar^2}\,V(r)
=-2(\beta^2-\alpha^2)\:\left(\cosh(\beta r) -
\frac{\alpha}{\beta}\sinh(\beta r)\right)^{\!-2}\:,
\end{equation}
with $\beta > \alpha >0$,
has precisely one bound state with $E=-\hbar^2\alpha^2/2m$.

Taking the ratio of the scattering to bound-state wave functions at low 
energies and distances then gives 
\begin{eqnarray}
\label{3_6}
\nonumber
R_1(r) &=& \frac{1}{2(\beta^2-\alpha^2)} 
- \frac{1}{6}\,r^2 - \frac{1}{90}\,(2\beta^2 -3\alpha^2)\,r^4+
0\!\left(r^6\right)\:,\\
R_2(r) &=& -\frac{1}{8(\beta^2-\alpha^2)^2}
-\frac{1}{12(\beta^2-\alpha^2)}\,r^2
-\frac{(\beta^2-3\alpha^2)}{360(\beta^2-\alpha^2)}\,r^4
+0\!\left(r^6\right). 
\end{eqnarray}
\vspace{2mm}

\noindent
{\large\bf C.\ \ The spherical well potential}

The solutions for the spherical well potential $U(r)=-U_{0}\,\theta(a-r)$,
discussed numerically in \S2, also lead to analytic expressions for the ratio
function. Thus
\begin{equation}
\label{3_7}
R_1(r) = \frac{a^2}{4(1+\alpha a)}-\frac{(4+\alpha a)}{4(U_0-\alpha^2)}
- \frac{1}{6}\,r^2 + 0\!\left(r^4\right)\:,
\end{equation}
where the binding energy $E=-\hbar^2\alpha^2/2m$ is determined from the 
implicit equation
\begin{equation}
(U_0-\alpha^2)^{1/2}\: \cot\left(a(U_0-\alpha^2)^{1/2}\right)=-\alpha\:.
\label{3_8}
\end{equation}
\vspace{2mm}

\noindent
{\large\bf D.\ \ Conclusions}

It is clear from the examples given above that for a lightly bound state
($\alpha \ll U(0))$, the expansion parameter at the origin is
proportional to $(\alpha^2+k^2)$
times the square of the range of the potential, though the coefficient 
depends upon the shape of the potential. However the $r$-dependence of the
expansion coefficients $R_n(r)$ can be obtained directly by expanding the
potential in powers of $r$ and integrating the Schr\"{o}dinger
equation starting from 
$r=0$. It is therefore no accident that the coefficient of $r^2$ in $R_{1}(r)$
is the same for the Bargmann case of Eq.(\ref{3_6}) as for the square well of
Eq.(\ref{3_7}). This result is independent of the potential provided 
that this is
finite at the origin. In the Yamaguchi example of Eq.(\ref{3_4}) the odd powers
of $r$ arise from the divergence of the potential at short distances.

It was noted in \S2 that in the case of the first lightly bound state in a
spherical well potential the scattering function lay above the bound-state
function, whereas if it were the second bound state then the scattering 
function lay below. This change in sign of $R_1(0)$ in Eq.(\ref{3_7}) arises
because in the first case $U_0\approx \pi^2/4a^2$, whereas in the 
second $U_0\approx 9\pi^2/4a^2$.

In the numerical examples shown in \S2, where the deviations from the
extrapolation theorem are not too large, a cross-over region was noted where
all the functions are very close. That the scattering function cross the
bound state, in a region where the latter wave function is large, follows
immediately from the condition that the two be orthogonal when integrated over
$r$. This position will be stable with respect to changes in $k$ provided that
the first term in the power series expansion in $(\alpha^2+k^2)$ in 
Eq.(\ref{3_2}) is dominant and
in such an event the cross-over point would be determined by the condition that
$R_1(r)=0$. For $\beta=1.0$ and $\alpha=0.1$ in the Bargmann case one would
expect that at low energies the cross-over would occur when $r\approx 1.74$ if
we keep only the quadratic term in $r^2$ in Eq.(\ref{3_6}), but $r\approx 1.68$
with the quartic term. These are to be compared to the numerical value of
$r\approx 1.522$. Of course the cross-over point should move to smaller values
of $r$ as the energy is increased since the scattering wave functions start to
oscillate faster and the node in $r$ is shifted to the left.

The above arguments do not provide any explanation as to the sign of $R_1(0)$
and for this the spherical well case is particularly illuminating. In the case
of fig.~1, where one has only one (lightly)-bound state, then $R_1(0)$ is
positive. This is also true for the other two potentials discussed in this
section. However when the depth of the well is increased so as to make the
lightly bound state the {\it second} one then $R_1(0)$ changes sign. This is 
in the right direction to make the extrapolation theorem of Eq.(\ref{1_6}) be 
valid also for the deeply bound state at more negative values of $k^2$
where $R(k,r)$ has to diverge to $+\infty$. On the 
other hand, when there is only one bound state the only singularities that
$R(k,r)$ can be simulating are those associated with the potential 
itself.\newpage

\noindent
{\large\bf IV.\ \ The Coulomb potential}\\

Due to the long range of the Coulomb potential the infinite number of 
bound-state eigenvalues have an accumulation point at energy zero. 
Nevertheless it is easy to see explicitly that the extrapolation 
theorem is still valid for all the bound states. 

Define full scattering wave functions as $\psi(k,r)=v(k,r)/r$, and similarly 
for the bound state. The value of the square of the {\it real} Coulomb wave 
function at the origin in an $S$-wave scattering state is just the Gamow 
factor 
\begin{equation}
[\psi(k,0)]^2 = \frac{2\pi\eta}{\mbox{\rm e}^{2\pi\eta}-1}
= \pi\eta \left[-1 + \coth (\pi\eta)\right]\,,
\label{4_1}
\end{equation}
where in the attractive case the Coulomb parameter $\eta=-m e^2/\hbar^{2}k$.

The infinite number of bound states correspond to the poles of Eq.(\ref{4_1})
when $\eta=in$, where $n$ a non-vanishing integer$^{8}$. These can be made 
explicit by recasting the equation as
\begin{equation}
\label{4_2}
[\psi(k,0)]^2 = - \pi\eta + 1+2\sum_{n=1}^{\infty}
\frac{1}{1+n^2/\eta^2} \,. 
\end{equation}

If we write the virtual momenta of the bound states as 
$\alpha_n= m e^2/n\hbar^2$,
it is well known that the value of the $n$'th bound-state wave function 
at the origin is
\begin{equation}
\label{4_3}
[\psi_n(0)]^2=4\alpha_n^3 \,,
\end{equation}
so that Eq.(\ref{4_2}) may be written as 
\begin{equation}
\label{4_4}
[\psi(k,0)]^2= \frac{\pi m e^2}{\hbar^{2}k} + 1 + \sum_{n=1}^{\infty}\,
\frac{[\psi_n(0)]^2}{2\alpha_n(\alpha_n^2+k^2)} \,\cdot
\end{equation}

Our theorem of Eq.(\ref{1_6}) is still formally valid since, in the vicinity of
any of the Coulomb bound states, just one term in the sum will be important and
for this the residue is clearly correct. 
%It is also easy to generalise 
%Eq.(\ref{4_4}) to the case of higher partial waves. 
Despite this, it is of 
no practical use in representing the scattering wave function at low 
energies since no single pole will then dominate and, as can be seen from 
Eq.(\ref{4_4}), the Coulomb potential produces an extra non-trivial term 
singular at $k=0$. Thus the expansion of $R(k,r)$ in powers of 
$(\alpha^2+k^2)$ as in Eq.(\ref{3_2}) is not valid. \\[3ex]

\noindent
{\large\bf V.\ \ The two-potential formula}\\

We here show the relation of our approximation to some of the results of 
standard perturbation theory. Denote by $f_0$ the S-wave scattering 
amplitude corresponding to a potential $V_0$. If a small extra potential $V_1$ 
is added, the change in the amplitude to first order in $V_1$ is given by 
the two-potential formula$^{3}$
\begin{equation}
\Delta f \approx-\frac{2m}{\hbar^2}\,\mbox{\rm e}^{2i\delta_0}\, 
\langle \chi_k|V_1| \chi_k\rangle
\,.
\label{5_1}
\end{equation}
The factor involving the phase shift $\delta_0$ for the potential $V_0$
is introduced because $\chi_k$ is the real scattering solution. Note that
the integration in the expectation value is only over the radial coordinate
$r$.

Assume now the potential $V_1$ to be of short range such that the conditions 
leading to Eq.(\ref{1_6}) are applicable. If the potential $V_0$ 
has one weakly bound state $\chi_b $, then $\chi_k$ is related to it through 
the approximate relation of Eq.(\ref{1_7}) and this leads to 
\begin{equation}
\Delta f\approx -\,\mbox{\rm e}^{2i\delta_0}\,
\frac{m}{\alpha_0(\alpha_0^2+k^2)
\hbar^2}\, \langle \chi_b |V_1| \chi_b \rangle 
=-\,\mbox{\rm e}^{2i\delta_0}\, \frac{m\,\Delta E}
{\alpha_0(\alpha_0^2+k^2)\hbar^2} \,\cdot
\label{5_2}
\end{equation}
In the above we have used perturbation theory to identify $\Delta E$ as the 
change in the binding energy $E_0=-\alpha_0^2/2m\hbar^2$ of the bound 
state to first order in $V_1$. In this limit
\begin{equation}
\mbox{\rm e}^{2i\delta} \approx 
1+2ik(f_0+\Delta f)=
\mbox{\rm e}^{2i\delta_0} 
\left[1-\frac{2ikm\triangle E}
{\alpha_0(\alpha_0^2+k^2)\hbar^2} \right] \,\cdot \label{5_3}
\end{equation}

For a loosely bound state, the S-matrix is given by the scattering length 
approximation 
\begin{equation}
\label{5_4}
 \mbox{\rm e}^{2i\delta_0} = \frac{\alpha_0-ik}{\alpha_0+ik}\ \ \ \
\mbox{\rm and}\ \ \ \ 
 \mbox{\rm e}^{2i\delta} = \frac{\alpha-ik}{\alpha+ik}\:\cdot
\end{equation}

The correction in Eq.(\ref{5_3}) is valid only to leading order in 
$\triangle E$ and introducing this into Eq.(\ref{5_4}) and keeping only such
terms, we find
\begin{equation}
\label{5_5}
 \alpha = \alpha_0\left( 1-\frac{m\triangle E}{\alpha_0^{\,2}\hbar^2}\right) 
\end{equation}
and a binding energy of
\begin{equation}
\label{5_6}
 E= -\frac{\hbar^2\alpha^2}{2m} =E_0+\triangle E\,,
\end{equation}
as expected. This explicitly shows the consistency of first order perturbation
theory for bound-state energies and scattering matrices with our approximate
wave function in the case of a weakly bound state.\\[3ex]
%-------------------------------------------------------------------

\noindent
{\large\bf VI.\ \ Summary and conclusions}\\

We have shown for several soluble examples that it is possible to approximate
the real scattering wave function quantitatively in terms of that of a nearby
bound state. This goes further than the theorem of Eq.(\ref{1_6}) in that it 
shows that it is a {\it robust} extrapolation provided that the binding is weak
and that the wave number $k$ and distance $r$ are both small. Though we have
worked entirely with real wave functions, this is sufficient for
applications in final-state-interaction theory$^{4}$. 

Provided that the bound-state pole is close, one can use 
Eq.(\ref{1_4}) to obtain the complex scattering wave function with 
outgoing boundary conditions in the scattering length approximation
\begin{equation}
\label{6_1}
\psi^{(+)}(k,r) =\psi(k,r)\,\mbox{\rm e}^{i\delta(k)} \approx
-(2\alpha)^{-1/2}\:\psi_{\alpha}(r)/(\alpha+ik)\:.
\end{equation}

Though we have a qualitative explanation for the deviations from the 
approximate relationship of Eq.(\ref{1_7}), its magnitude at 
$r=0$ must depend upon the structure of the potential and the 
singularities of the scattering amplitude. Fortunately the cross-over 
phenomenon, whereby all the functions come very close at some region 
within the range of the potential, limits the deviations significantly.
As explained in \S3, this cross-over phenomenon is very general and exists 
also for realistic nucleon-nucleon potentials, such as that of the Paris 
group$^{4,9}$. It is also demonstrated by numerical resolution of the 
Schr\"{o}dinger equation for various finite-range potentials.\\

\noindent
{\large\bf Acknowledgements}

We should like to thank Bengt Karlsson for innumerable discussions. This work
has been made possible by the continued financial support of the Swedish Royal
Academy and the Swedish Research Council, and one of the authors (CW) would 
like to thank them and the The Svedberg Laboratory for generous 
hospitality.\\[3ex]

\noindent
{\large\bf References}\\
\begin{tabbing}
$^{a}$ \=Electronic address: faldt@tsl.uu.se\\
$^{b}$ \>Electronic address: cw@hep.ucl.ac.uk\\
$^{1}$ \>G.~F\"aldt and C.~Wilkin, ``Scattering wave functions at bound
state poles'',\\ \>Physica Scripta {\bf 56}, 566 (1997).\\
$^{2}$ \>L.I.~Schiff, {\it Quantum Mechanics} (McGraw-Hill, New York,
1968), 3rd ed., p.349.\\
$^{3}$ \>C.J.~Joachain, {\it Quantum Collision Theory} (North Holland,
Amsterdam, 1975),\\ \>pp.\ 254-258, and p.\ 449.\\
$^{4}$ \>G.F\"aldt and C.Wilkin, ``A comparison of $np\to d\eta$ and
$np\to np\eta$ production\\ \>rates'', Nucl.\ Phys.\ A{\bf 604}, 441-454
(1996).\\
$^{5}$  \>G.F\"aldt and C.Wilkin, ``Bound state and continuum production
in large \\ \>momentum transfer reactions'', Phys.\ Lett.\ B{\bf 382},
209-213 (1996)\\
$^{6}$  \>G.F\"aldt and C.Wilkin,
``Cross section and analyzing power of $\vec{p}p\to pn\pi^+$
\\ \>near threshold'', Phys.Rev. C{\bf 56}, 2067-75 (1997).\\
$^{7}$ \>A.Boudard, G.F\"aldt and C.Wilkin, ``Triplet $np$ final state
interactions at large \\ \>momentum transfers'', Phys.\ Lett.\ B{\bf 389}, 
440-444 (1996).\\
$^{8}$ \>R.H.~Landau, {\it Quantum Mechanics II}, (Wiley, New York,
1990), 2nd ed., p.~60.\\
$^{9}$ \>M.~Lacombe {\it et al.}, ``Parametrization of the Paris $N-N$
potential'', \\ \>Phys.Rev. C{\bf 21}(3) 861-873 (1980).
\end{tabbing}

\input epsf
\begin{figure}
\vspace*{-3cm}
\begin{center}
\mbox{\epsfxsize=7cm \epsfbox{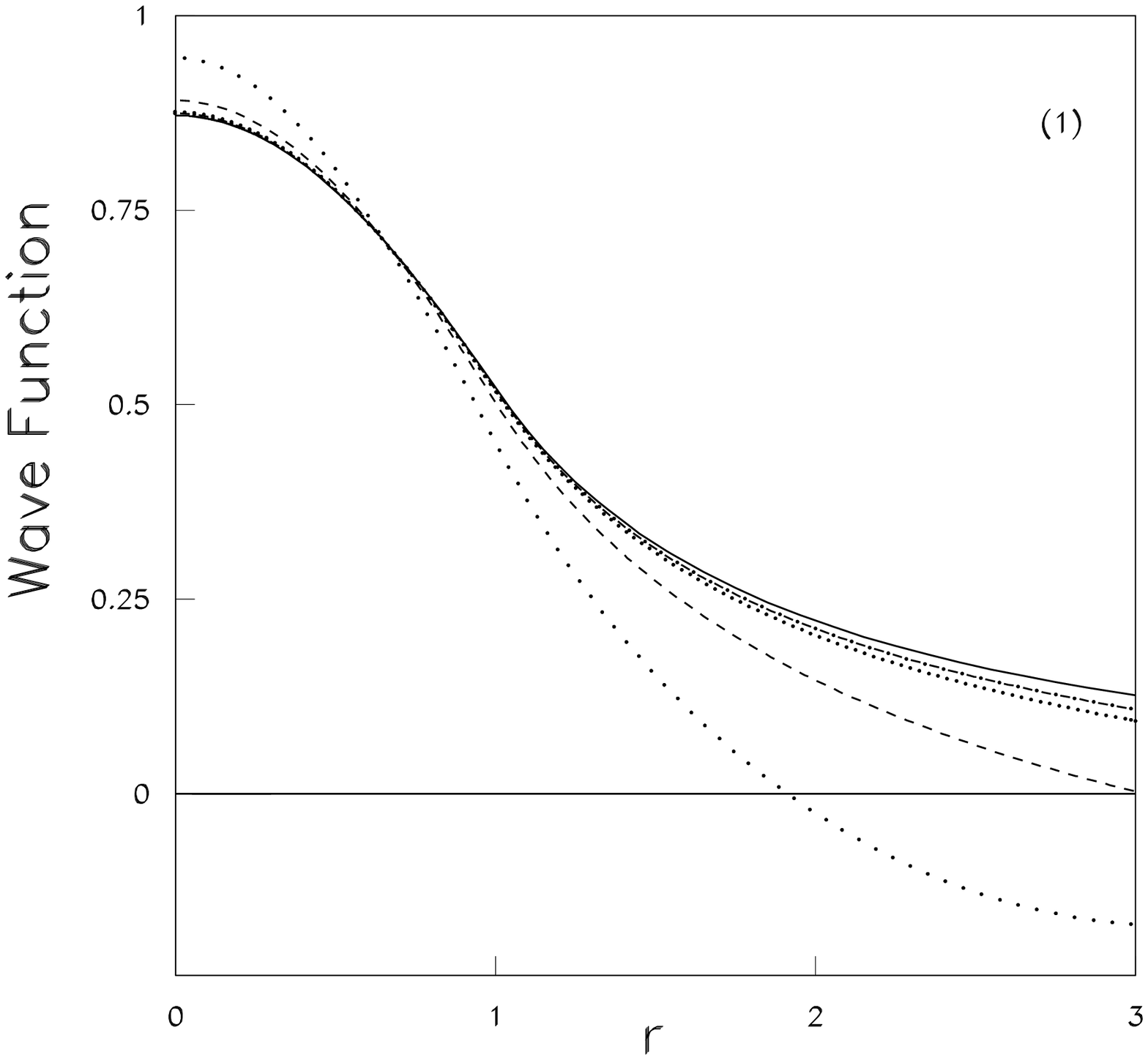}}
\mbox{\epsfxsize=7cm \epsfbox{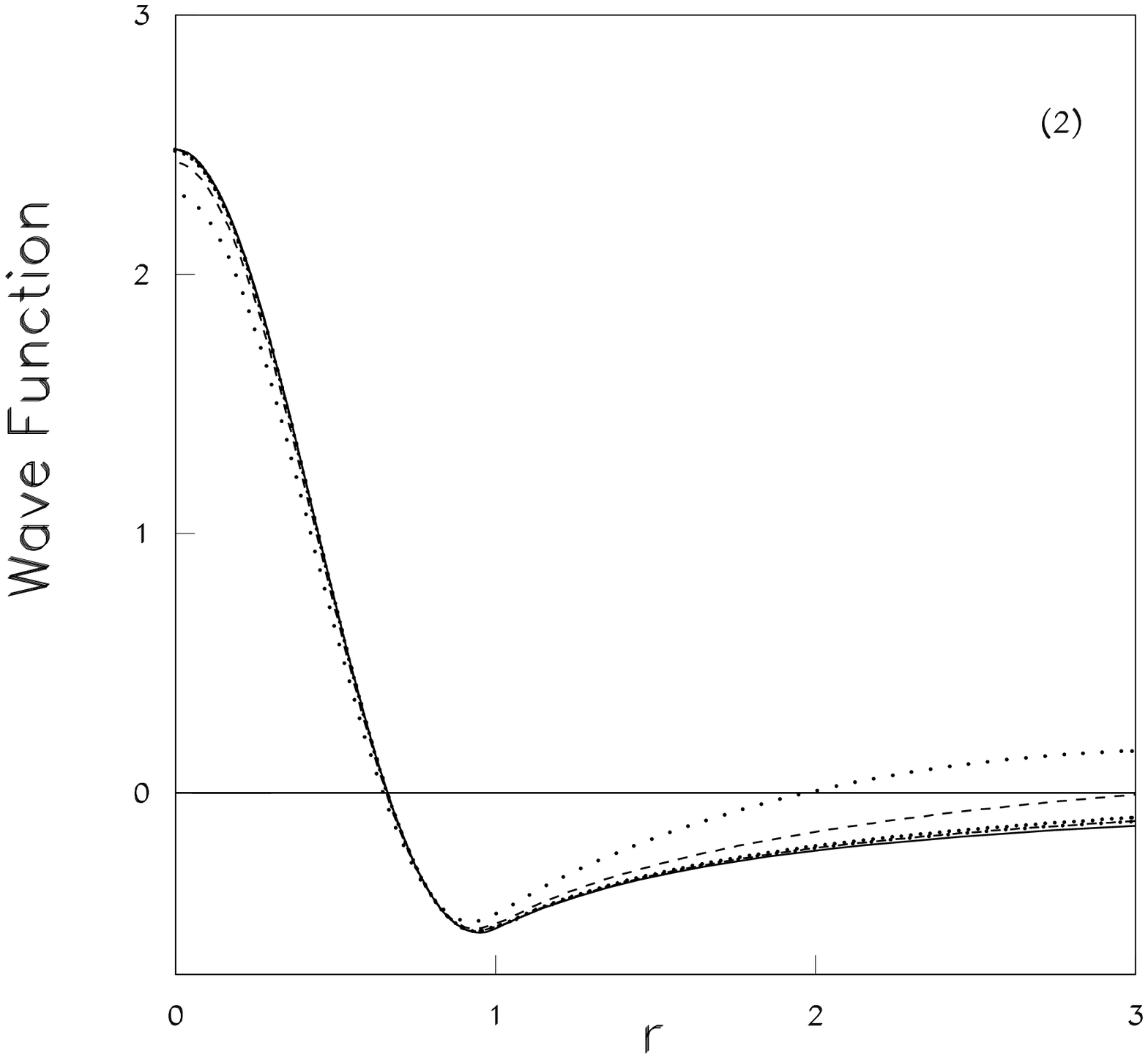}}
\mbox{\epsfxsize=7cm \epsfbox{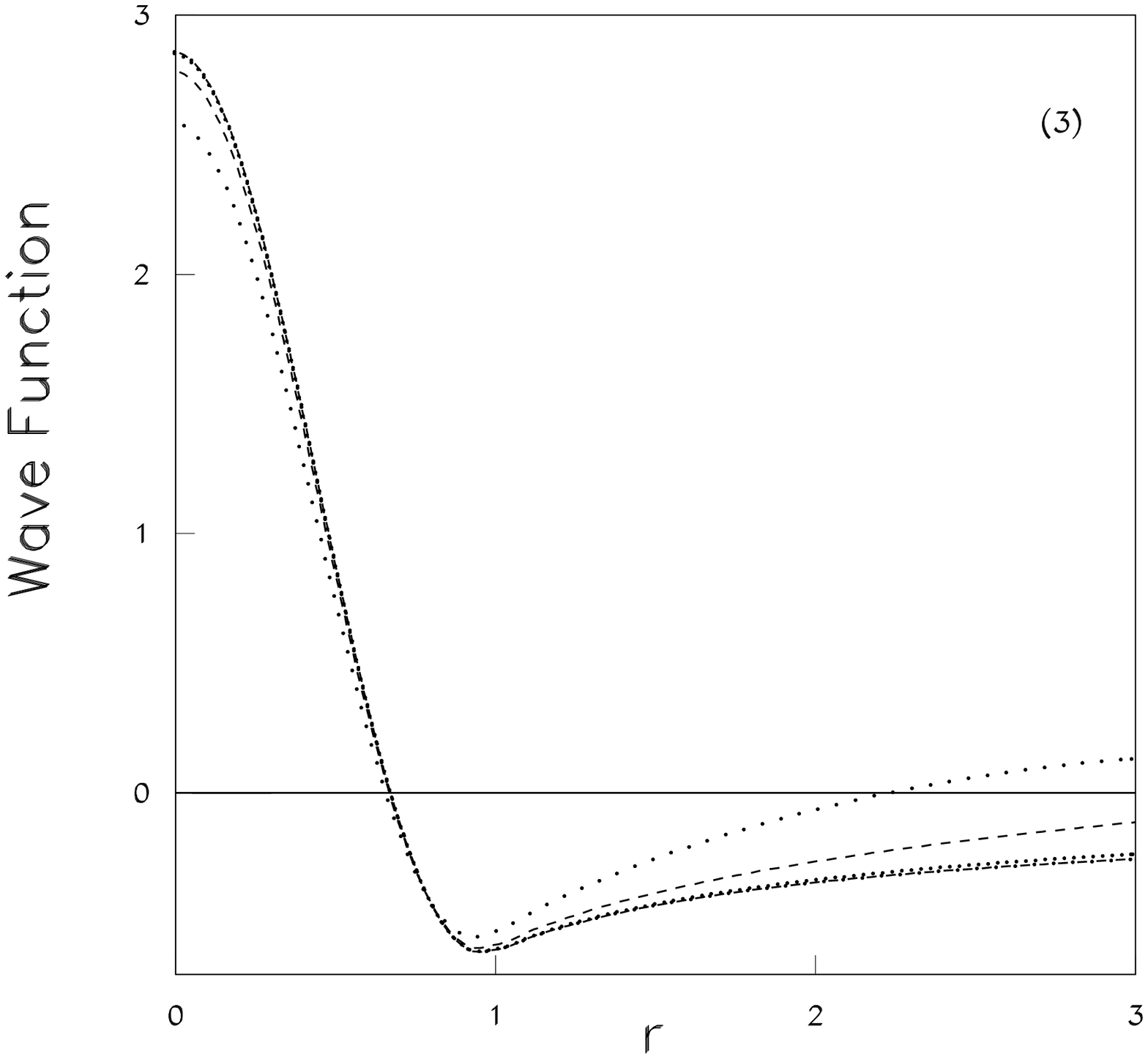}}
\caption{\baselineskip 4ex
Bound-state (solid line) and scattering wave functions, modified 
as in Eq.(\protect\ref{2_2}), for a spherical well of radius $1.0$. 
The scattering states are evaluated at $k=0.1$ (dot-dashed line), $k=0.2$
(closely spaced dots), $k=0.5$ (dashes), and $k=1.0$ (widely spaced dots).
(1) For a depth of $U_0=2.8$, the resulting $1s$ bound state has 
$\alpha=0.159$, and the deviations at $r=0$ are positive and increase 
with $k$.
(2) A potential of strength $U_0=22.547$ has a $2s$ level with the same
value of $\alpha=0.159$. The scattering functions now lie below that
of the bound state at $r=0$.
(3) Reducing the potential strength to $U_0=21.913$ turns the $2s$
level into a virtual state with $\alpha=-0.159$. Nevertheless the scattering
functions retain their dominantly $2s$ character at short distances and
illustrate that it is the nearby singularity, bound or virtual, which governs
the energy dependence of the scattering wave functions.}
\end{center}
\end{figure}
\end{document}